\documentstyle[preprint,tighten,epsf,aps]{revtex}
\newcommand{\mafigura}[4]{
\begin{figure}[hbtp]
\begin{center}
\epsfxsize=#1 \leavevmode \epsffile{#2}
\end{center}
\caption{#3}
\label{#4}
\end{figure} }
\newcommand{\eq}[1]{eq.~(\ref{#1})}
\newcommand{\as}{\alpha_s}

\newcommand{\ra}{\rightarrow}
\newcommand{\beq}{\begin{equation}}
\newcommand{\eeq}{\end{equation}}
\newcommand{\bea}{\begin{eqnarray}}
\newcommand{\eea}{\end{eqnarray}}
\newcommand{\non}{\nonumber}
\begin{document} 


\title{$m_b(m_Z)$ from jet production at the $Z$ peak\\
in the Cambridge algorithm} 


\author{Mikhail Bilenky\thanks{On leave from JINR, 141980 Dubna, Russian Federation}}
\address{Institute of Physics, AS CR, 18040 Prague 8, and 
Nuclear Physics Institute, 
AS CR, 25068 \v{R}e\v{z}(Prague), Czech Republic}
\author{Susana Cabrera and Joan Fuster}
\address{IFIC, CSIC-Universitat de Val\`encia, 46100 Burjassot, 
Val\`encia, Spain} 
\author{Salvador Mart\'{\i}}
\address{CERN, European Laboratory for Particle Physics, CH-1211 Gen\`eve, Switzerland}
\author{Germ\'an Rodrigo}
\address{INFN - Sezione di Firenze, Largo E. Fermi 2, 50125 Firenze, Italy}
\author{Arcadi Santamaria}
\address{Departament de F\'{\i}sica Te\`orica, IFIC,
CSIC-Univ. de Val\`encia, 46100 Burjassot, Val\`encia, Spain} 
\maketitle 

\begin{abstract}
We consider the production of heavy quark jets at the $Z$-pole at the
next-to-leading order (NLO) using the {\it Cambridge jet-algorithm}.
We study the effects of the quark mass in two- and three-jet observables
and the uncertainty due to unknown higher order corrections as well as
due to fragmentation. We found that the three-jet observable has remarkably
small NLO corrections, which are stable with respect to the change of the
renormalization scale, when expressed in terms of the
{\it running quark mass} at the $m_Z$-scale.  The size of the hadronization
uncertainty for this observable remains reasonably small and is very stable
with respect to changes in the jet resolution parameter $y_c$. 
\end{abstract}


\section{Introduction} 

During last few years a significant progress has been done in 
the understanding of the heavy quark jet-production in $e^+e^-$-annihilation
both experimentally~\footnote{See~\cite{Burrows:1998ab} for a review of
recent experimental results.} and theoretically. 

The DELPHI collaboration has measured the bottom-quark
mass~\cite{MartiiGracia:1997ak,Abreu:1997ey} analyzing the
$e^+e^-$-annihilation into the three-jet final state with heavy quarks 
using recent next-to-leading order theoretical predictions
for this process~\cite{Rodrigo:1997gy,Rodrigo:1997gv,Rodrigo:1996gw,Brandenburg:1997pu,Bernreuther:1997jn,Oleari:1997az,Nason:1997nw}.
The DELPHI result~\footnote{The $\overline{MS}$ definition for the
running mass at the $m_Z$-scale was used.}
for the Durham~\cite{Catani:1991hj} jet clustering algorithm 
\beq
m_b(m_Z)=2.67\pm0.25(stat.)\pm0.34(had.)\pm0.27(theo.) GeV~, 
\label{eq:delphi_mb}
\eeq
was the first measurement of the $b$-quark mass far above the
production threshold and it is the first experimental evidence
(at the 2-3 sigma level) of the running of a fermion mass,
as predicted by the Standard Model.
Recently, the SLD collaboration has also analyzed its three- and four-jet 
data using Durham and several Jade-like jet algorithms~\cite{sld}.
The value of the b-quark mass~\cite{Brandenburg:1999nb} obtained from these 
data is compatible with the above DELPHI result.

The heavy quark mass measurement was done under the assumption of
the flavor independence of the strong interactions with the value
of the strong coupling constant, $\as$,
fixed to its world average measured in other experiments.
On the other hand, assuming a given $b$-quark mass value obtained from
low-energy measurements, and comparing the value of 
$\as$ measured from the heavy quark three-jet final state
with the one measured from the production of light quarks,
one can perform a test of the flavor universality of the strong
interaction. Such a test was performed
recently~\cite{Abreu:1997ey,sld,opal} and no deviation from the QCD
prediction was found. The next-to-leading order QCD predictions with
heavy quark mass corrections 
from references~\cite{Rodrigo:1997gy,Rodrigo:1997gv,Rodrigo:1996gw,Brandenburg:1997pu,Bernreuther:1997jn,Oleari:1997az,Nason:1997nw}
were used in these studies.

There are three main sources of uncertainties in the DELPHI analysis.
The first one has statistical nature. The second error is
due to the uncertainty in the hadronization corrections. It was 
evaluated~\cite{Abreu:1997ey} using different Monte-Carlo models
simulating the hadronization process.
The third one is due to our ignorance of higher order perturbative
corrections in the theoretical predictions at the partonic level.
The last uncertainty was estimated by varying the renormalization
scale in the calculations and by using different renormalization
schemes, i.e. expressing intermediate results in terms of either
the perturbative pole quark mass or the running quark mass. 

The value  of the $b$-quark mass measured at the
$Z$ peak (\ref{eq:delphi_mb}) is found to be in good agreement with the 
determinations of
the $b$-quark mass at low-energy from $\Upsilon$- and $B$-mesons
spectroscopy~\cite{mbmz}, when compared at the same scale.
However, the uncertainties in $m_b(m_Z)$ are larger.
Thus, it would be desirable to reduce this error by finding new
observables which may show a better theoretical and hadronization
properties.

In this paper we study quark mass effects in heavy quark jet
production by using the Cambridge jet clustering 
algorithm~\cite{Dokshitser:1997in,Moretti:1998qx}. We also study
the possibility to reduce the uncertainties in the measurement of the 
$b$-quark mass at the $Z$-pole.
We consider two jet observables and estimate the errors in their
theoretical predictions due to the unknown higher orders by varying
the renormalization scale and considering different renormalization
schemes. We discuss also the size of the uncertainty due to the
hadronization process.  Some preliminary results of this study
were reported in~\cite{Rodrigo:1998vq,juan98}.

\section{The Cambridge algorithm and the decay $Z \ra 3jets$ 
at the next-to-leading order} 

The Cambridge~\cite{Dokshitser:1997in} jet clustering algorithm
is a modified version of the popular Durham~\cite{Catani:1991hj}
algorithm that has been introduced recently in order to reduce
at low $y_c$ the formation of spurious jets with low transverse
momentum particles. Consequently, compared to Durham,
it allows to explore regions of smaller $y_c$ while still keeping
higher order corrections relatively small. It is important to note that
at low $y_c$
the statistical experimental error for three-jet and four-jet
production is expected to be smaller, and the sensitivity to the
quark mass increases. 

In the Durham algorithm one finds the minimal test variable
$y_{ij}$ defined as
\beq
y_{ij} = 2 \frac{min(E_i^2,E_j^2)}{s}(1-\cos \theta_{ij})~,
\label{eq:testva}
\eeq
for all possible pair-combinations of the particles and compare
it with the jet-resolution parameter, $y_c$. In \eq{eq:testva} 
$E_i$ and $E_j$ denote the energies of particles~\footnote{By the word
"particles" we mean here both the real hadrons detected at experiment
and the partons entering the theoretical calculation.} $i$ and $j$,
$\theta_{ij}$ is the angle between their three-momenta and $s$ is
the center-of-mass energy squared. 
If $y_{ij}<y_c$, the two particles $i$ and $j$ are combined into a
new pseudoparticle with momentum
\beq
p_k=p_i+p_j~.
\label{eq:combi}
\eeq
The procedure is repeated again and again until $y_{ij}>y_c$ for all
pairs of (pseudo)particles. The number of (pseudo)particles 
at the end defines the number of jets.

The Cambridge algorithm is defined by the same test variable,
\eq{eq:testva}, and the same recombination rule, \eq{eq:combi},
as Durham. The new ingredient of the Cambridge algorithm is the 
so-called {\it ordering variable} 
\beq
v_{ij} = 2 (1-\cos \theta_{ij})~.
\eeq
In this algorithm, first, the pair of particles, which has minimal 
ordering variable $v_{ij}$, is selected. Then one computes $y_{ij}$
for this pair of particles, and, if $y_{ij}<y_c$, the two particles
are recombined into a pseudoparticle according to~\eq{eq:combi}.
But if $y_{ij}>y_c$, the softer particle from this pair is assigned
to a resolved jet. This last step is called "soft-freezing". 

Because of the additional step in the jet finding iterative procedure, 
the Cambridge scheme turns out to be more complex and has a number of
peculiar properties~\cite{Bentvelsen:1998ug}.
Let us mention only one example.
In Jade-like algorithms, including Durham, one can always define a
transition value of $y_c$, such that a multi-parton event classified
as a $n$-jet event becomes a $n+1$-jet event, when the value of $y_c$
is slightly decreased. However, as pointed out in~\cite{Bentvelsen:1998ug},
this property is lost in the Cambridge algorithm, since, due to the presence
of the ordering parameter, the sequence of clustering depends on the
value of $y_c$. As a result the number of jets is not a monotonic
function of $y_c$ and it can change by a non unit number at some
transition value of $y_c$. 

With the above definitions one can show that the cross section of
the $e^+e^-$-annihilation into three jets calculated at the leading
order (LO) is the same in both Cambridge and Durham algorithms.
This happens because only three-parton final state configurations 
contribute at LO to the three-jet cross section. Instead, the
four-jet production cross section at the LO is different in the
two schemes.

At the next-to-leading order (NLO) the predictions for the three-jet
production cross section for the two algorithms are different.
Schematically, the NLO calculation of $e^+e^- \ra 3jets$ was performed
as follows. In this case the three-jet cross section receives contributions
not only from one-loop corrected three-parton final states, but also from
four-parton processes. In the latter process two of the four partons
are combined in order to produce a three-jet final state.
The ultraviolet (UV) divergences encountered in the calculation
of the three-parton contribution at the one-loop level
were removed by the renormalization of the parameters of the QCD Lagrangian.
The infrared (IR) divergences~\footnote{Dimensional regularization 
was used to regularize both ultraviolet (UV) and infrared (IR) singularities 
in the whole calculation.} remaining in this part, which are due to the
presence of massless gluons in the loop, were canceled in the final
result for the three-jet transition probability by adding an appropriate
contribution from the four-parton final state. 
In the latter contribution, which is a purely tree-level one,
the IR divergences appear due to the radiation of soft or/and collinear
massless gluons. To separate the IR divergent part of the four-parton
contribution, the phase-space slicing method (see~\cite{Keller:1999tf}
and references cited therein) has been used.
In this method the integration over a thin slice at the edge of the
phase-space (containing the soft and the collinear singularities)
is performed analytically. Then, the IR singularities coming from
three- and four-parton final states are canceled analytically.
The remaining finite pieces from both three-parton and four-parton
processes are integrated numerically over the three-jet phase-space
defined by the specific jet algorithm. The four-jet cross section at
the leading order, which is IR finite, is also obtained by numerical
integration over the four-jet part of the four-parton phase-space.

Details of the NLO calculation for the Durham and some
other popular jet clustering algorithms were presented
in~\cite{Rodrigo:1997gy,Rodrigo:1997gv,Rodrigo:1996gw}. 
In the case of the Cambridge algorithm, although all principal
calculational steps of the three-jet heavy quark production in
$e^+e^-$-annihilation at NLO remain the same as in other algorithms,
the practical implementation of this scheme turned out to be more
involved due to more complex realization of the Cambridge jet finder. 

\section{The Observables} 

In this paper we study in detail the following
ratio of three-jet rates in the Cambridge jet-algorithm
\beq
R_3^{b\ell} = \frac{\Gamma_{3j}^b(y_c)/\Gamma^b}
{\Gamma_{3j}^\ell(y_c)/\Gamma^\ell}~.
\label{eq:r3bl}
\eeq
In the above equation $\Gamma_{3j}^b$ and $\Gamma^b$ stand, respectively,
for the three-jet and the total decay widths of the $Z$-boson
with a $b$-quark in the final state. 
Analogously, the quantities with the superscript $\ell$ denote
the sum of the decay widths into light quarks ($\ell=u,d,s$) 
which all are considered massless.

The ratio $R_3^{b\ell}$ can be written
in the form of the following expansion in $\alpha_s$ 
\beq
R_3^{b\ell} = 1 + \frac{\alpha_s(\mu)}{\pi} a_0(y_c) + r_b \left( b_0(r_b,y_c) + 
\frac{\alpha_s(\mu)}{\pi} \: b_1(r_b,y_c) \right)~,
\label{eq:pole}
\eeq
where $r_b = M_b^2/s$, with $M_b$ the heavy quark pole mass,
and $s=m_Z^2$ at the $Z$ peak.

Let us remark that the double ratio in~\eq{eq:r3bl} differs slightly
from the one, $R_3^{bd}$, considered in~\cite{Bilenkii:1995ad,Rodrigo:1997gy}
with a normalization to the $Z$-decay width of only one light flavor,
the $d$-quark. In contrast to $R_3^{b\ell}$, such double ratio is equal
to unit for a vanishing $b$-quark mass.
The main difference between the two observables, $R_3^{b\ell}$ and $R_3^{bd}$,
is due to the triangle one-loop diagrams~\cite{Hagiwara:1991dx},
which give a non-zero contribution even in the case of massless
$b$-quarks taken into account by a function $a_0(y_c)$ in \eq{eq:pole}.
The difference is, however, very small numerically, smaller than a $0.2$\%.

The functions $b_0$ and $b_1$ in~\eq{eq:pole} describe, accordingly,
the quark mass effects at the leading and the next-to-leading order
in the strong coupling and depend on the jet clustering scheme.
Although, for convenience, the leading polynomial dependence 
on $r_b$ has been factorized out in~\eq{eq:pole}, the exact dependence 
on the heavy quark mass is kept in the functions $b_0(r_b,y_c)$ and
$b_1(r_b,y_c)$.

Using the known relationship between the perturbative pole mass
and the $\overline{MS}$ scheme running mass~\cite{Tarrach:1981up},
\beq
M_b^2 = m_b^2(\mu) \left[1+\frac{2\alpha_s(\mu)}{\pi} 
\left(\frac{4}{3} -\log \frac{m_b^2}{\mu^2} \right)\right]~,
\label{eq:poletorunning}
\eeq
we can re-express~\eq{eq:pole} in terms of the running mass
$m_b(\mu)$. Then, keeping only terms of order ${\cal O}(\alpha_s)$
we obtain
\beq
R_3^{b\ell} = 1 + \frac{\alpha_s(\mu)}{\pi} a_0(y_c) 
+ \bar{r}_b(\mu) \left( b_0(\bar{r}_b,y_c) + 
\frac{\alpha_s(\mu)}{\pi} \bar{b}_1(\bar{r}_b,y_c,\mu) \right)~,
\label{eq:MS}
\eeq
where $\bar{r}_b(\mu)=m_b^2(\mu)/m_Z^2$ and 
the new function $\bar{b}_1$ is related to $b_0$ and $b_1$,
introduced in~\eq{eq:pole}, via
\beq
\bar{b}_1(\bar{r}_b,y_c,\mu) =
b_1(\bar{r}_b,y_c) + 2 b_0(\bar{r}_b,y_c) 
\left(\frac{4}{3} - \log \bar{r}_b + \log \frac{\mu^2}{m_Z^2} \right)~.
\label{eq:b1bar}
\eeq
Effectively, the use of $m_b(\mu)$, instead of $M_b$, corresponds 
to the use of a different renormalization scheme. Although at the
perturbative level both expressions, \eq{eq:pole} and \eq{eq:MS},
are equivalent, numerically they give different answers since
different higher order contributions are neglected. 
The spread of the results gives an estimate of the size of higher
order corrections.

As discussed in the previous section, the phase-space integration
(up to five-fold) in the calculation of the NLO decay width of
the $Z$-boson into three jets is done numerically. This numerical
integration is rather time consuming. Hence, we found it
very convenient to fit the numerical results with relatively
simple analytical functions. 
Because very small $y_c$-values are considered in the case of the
Cambridge scheme, these fits are more complex and involve more
parameters than the ones for the Durham algorithm described
in~\cite{Rodrigo:1997gy}. 
A Fortran code containing the fits to the functions
$b_0$ and $b_1$ (or $\bar{b}_1$)\footnote{Although, the two
functions $b_1$ and $\bar{b}_1$ are related via~\eq{eq:b1bar}
we performed independent fits for these functions.}
can be obtained from the authors upon request. The numerical
results for $R_3^{b\ell}$ are presented in the next section.

In the next section we also give numerical results for
the ratio of differential two-jet rates, defined as follows
\beq
D_2^{b\ell} =
\frac{[\Gamma_{2j}^b(y_c+\Delta y_c/2)
-\Gamma_{2j}^b(y_c-\Delta y_c/2)]/\Gamma^b}
{[\Gamma_{2j}^\ell(y_c+\Delta y_c/2)
-\Gamma_{2j}^\ell(y_c-\Delta y_c/2)]/\Gamma^\ell}~.
\label{eq:d2bl}
\eeq
Here, $\Gamma_{2j}^b$ and $\Gamma_{2j}^\ell$ denote the 
two-jet decay widths of the $Z$-boson with a $b$-quark
and light quarks in the final state, correspondingly.
The two-jet decay width at the order ${\cal O}(\alpha_s^2)$ is
calculated from the three- and the four-jet widths through the identity 
\[
\Gamma^q = \Gamma_{2j}^q + \Gamma_{3j}^q + \Gamma_{4j}^q~,
\]
where $q$ is the quark flavor and $\Gamma_{4j}^q$ is the 
four-jet decay width at the leading order. 
The value of $\Delta y_c $ in~\eq{eq:d2bl} should be chosen small enough.
We fix $\Delta y_c =0.001$ in the numerical analysis.

The differential ratio $D_2^{b\ell}$
is interesting because it contains different information than
the ratio $R_3^{b\ell}$. In addition, while
values of $R_3^{b\ell}$ measured at different $y_c$ are strongly correlated,
the differential rate $D_2^{b\ell}$ can be analyzed as a function of $y_c$.
The whole consideration of $R_3^{b\ell}$ discussed above is also
applied here. For $D_2^{b\ell}$ we use expansions in $\as$ similar to
those in~\eq{eq:pole} and~\eq{eq:MS}, see~\cite{Rodrigo:1997gv},
and we fit the corresponding LO and NLO numerical results to
simple analytical functions equivalent to $b_0$ and $b_1$ ($\bar{b}_1$).

\section{Perturbative results}

In fig.~\ref{fig:observables}  we present the results
for the two observables studied, $R_3^{b\ell}$ and $D_2^{b\ell}$,
as functions of the jet-resolution parameter $y_c$ in the
Cambridge algorithm. 
We plot the NLO results written either
in terms of the pole mass (\eq{eq:pole}) with 
$M_b=4.6$~GeV, or in terms of the running 
quark mass at $m_Z$ (\eq{eq:MS}) with  $m_b(m_Z)=2.8$~GeV.
The renormalization scale is fixed to $\mu=m_Z$ and
$\alpha_s(m_Z)=0.118$. For comparison we also show $R_3^{b\ell}$
and $D_2^{b\ell}$ at LO when the value of the pole mass, $M_b$, or
the running mass at $m_Z$, $m_b(m_Z)$, is used for the quark mass.
Note that one can not distinguish between different definitions
of the quark mass in the lower order calculation.
Mass effects monotonically grow for decreasing $y_c$,
they are very significant for both observables and in the case
of $D_2^{b\ell}$ exceed $10\%$ for small values of $y_c$.

 From this figure one sees a remarkable feature of the NLO result 
in the considered range of $y_c$, $0.005<y_c<0.025$, for the Cambridge
scheme : the NLO corrections, in the case when the running mass is used
are significantly smaller, especially for $R_3^{b\ell}$,
than the corrections in the case with the parameterization in terms
of the pole mass. In other words, using the running mass at the
$m_Z$-scale in the LO calculations
takes into account the bulk of the NLO corrections.
This situation, although does not guarantee, suggests that also 
next-to-next-to-leading and higher order corrections are small
for the observables parameterized in terms of the running mass
at the $m_Z$-scale, i.e. one has a better description of mass effects
in terms of a short distance parameter, $m_b(m_Z)$, than in 
terms of a low-energy parameter like the perturbative pole mass. 

The theoretical prediction for the observables studied contains a
residual dependence on the renormalization scale $\mu$:
when written in terms of the pole mass it only comes from the
$\mu$-dependence in $\alpha_s(\mu)$, when written in terms of
the running mass it comes from both $\alpha_s(\mu)$ and the
incomplete cancellation of the $\mu$-dependences between $m_b(\mu)$
and the logs of $\mu$ which appear in~\eq{eq:b1bar}.
The dependence on $\mu$ is usually regarded as an estimate of
the effect of the unknown higher order perturbative corrections.
In fig.~\ref{fig:muscales}.a we present the $\mu$-dependence of
the two NLO predictions, the pole mass prediction (NLO-$M_b$)
given by~\eq{eq:pole} and the running mass prediction (NLO-$m_b(m_Z)$)
given by~\eq{eq:MS}, for the ratio $R_3^{b\ell}$ 
in the range $m_Z/10 < \mu < m_Z$ at a fixed value of $y_c$.
We use the following one-loop evolution equations 
\beq 
a(\mu) = \frac{a(m_Z)}{K}~,
\qquad
m_b(\mu) = m_b(m_Z) \: K^{-\gamma_0/\beta_0}~,
\label{eq:mbrunning}
\eeq
where $a(\mu) = \as(\mu)/\pi$, 
$K = 1 + a(m_Z) \beta_0 \log (\mu^2/m_Z^2)$ and
\[
\beta_0 = \frac{1}{4} \left[11-\frac{2}{3}N_F\right]~, \qquad
\gamma_0 = 1~,
\]
with $N_F=5$ the number of active flavors, to obtain $\as(\mu)$ and
$m_b(\mu)$ from $\as(m_Z)=0.118$ and $m_b(m_Z)=2.8$~GeV.
The NLO-$m_b(m_Z)$ result (dashed line) shows a remarkable stability with
respect to the variation of the renormalization scale and the
corrections with respect to the LO prediction (LO-$m_b(m_Z)$)
remain small for all the values of $\mu$. 
Instead, the NLO-$M_b$ prediction (dotted line) has noticeably stronger
dependence on the renormalization scale. The NLO corrections in this case
remain sizable for all the values of $\mu$ and increase for decreasing $\mu$.
Note also that, as one would expect, for low values of $\mu$
the two NLO predictions, in terms of the running mass and in terms
of the pole mass, become very close to each other.

For a given value of $R_3^{b\ell}$ we can solve
\eq{eq:pole} (or \eq{eq:MS}) with respect to the quark mass.
The result, shown in fig.~\ref{fig:muscales}.b for a fixed value
of $R_3^{b\ell}$, depends on which equation was used and has
a residual dependence on the renormalization scale $\mu$.
The curves in fig.~\ref{fig:muscales}.b are obtained in the following way:
first from \eq{eq:MS} we directly obtain for an arbitrary value
of $\mu$ between $m_Z$ and $m_Z/10$ a value for the bottom-quark running
mass at that scale, $m_b(\mu)$, and then using \eq{eq:mbrunning}
we get a value for it at the $Z$-scale, $m_b(m_Z)$.
Second, using \eq{eq:pole} we extract, also for an arbitrary value
of $\mu$ between $m_Z$ and $m_Z/10$, a value for the
pole mass, $M_b$. Then we use \eq{eq:poletorunning} at $\mu=M_b$
and again \eq{eq:mbrunning} to perform the evolution from $\mu=M_b$
to $\mu=m_Z$ and finally get a value for $m_b(m_Z)$.
The two procedures, we denote them NLO-$m_b(m_Z)$ and NLO-$M_b$
respectively, give different answers, since different
higher orders have been neglected in the intermediate steps. 
The maximum spread of the two results in the whole $\mu$-range under
consideration can be interpreted as an estimate of the size of higher
order corrections, i.e. of the theoretical error in the determination
of the bottom-quark mass from the experimental measurement of $R_3^{b\ell}$.

We see from fig.~\ref{fig:muscales}.b that the first approach
is very stable with respect to the choice of the scale
used in \eq{eq:MS}. The obtained $b$-quark mass, $m_b(m_Z)$,
varies only $\pm 50$~MeV when the scale is varied in the 
range $\mu=m_Z$ and $\mu=m_Z/10$. 
In the same range of $\mu$, the estimated error in the Durham 
algorithm was found~\cite{Rodrigo:1997gy} to be $\pm 200$~MeV.
On the contrary, if one uses \eq{eq:pole}
the extracted quark mass has a strong scale dependence,
specially for small $\mu$-values and the estimated error
is very sensitive to the choice of the smallest possible value
of the renormalization scale.
Cutting as before at $\mu=m_Z/10$, the extracted pole mass
varies in the range $\pm 300$~MeV which is translated into
$\pm 240$~MeV for $m_b(m_Z)$. 
Let us note that a further $\pm 20$~MeV should be added due
to the uncertainty in the strong coupling constant 
$\Delta \alpha_s(m_Z) = \pm 0.003$.

Although, our observables are formally of order ${\cal O}(\alpha_s)$ 
and, therefore, compatible with the use of one-loop renormalization
group equations (RGE) to connect the running parameters at different 
scales, as a check of the stability of our results we have also repeated 
the analysis using two-loop evolution equations~\cite{Rodrigo:1998zd}
\bea
& & a(\mu) = \frac{a(m_Z)}
{K + a(m_Z) \: b_1 \:  \left( L + a(m_Z) b_1  
\frac{\displaystyle 1-K+L}{\displaystyle K} \right) }~, \non \\
& & m_b(m_Z) = m_b(\mu) \: K^{g_0} \:
\frac{1+a(m_Z) \: c_1}{1+a(\mu) \: c_1}~, 
\eea
where $L=\log K$ and $c_1=g_1-b_1 g_0$ with $b_1=\beta_1/\beta_0$,
$g_i=\gamma_i/\beta_0$ and 
\beq
\beta_1 = \frac{1}{16} \left[102-\frac{38}{3}N_F\right]~, \qquad
\gamma_1 = \frac{1}{16} \left[\frac{202}{3}-\frac{20}{9}N_F\right]~.
\eeq
The use of two-loop RGE corresponds to the dashed lines in
Fig~\ref{fig:muscales}.b.
Again a value of the quark mass extracted via the running mass 
parameterization remains more stable with respect to variation
of the scale $\mu$ and changes only slightly.
The mass extracted through the pole mass receives a significant
shift of $200$~MeV when the two-loop RGE are used.

\section{Hadronization corrections}

In the DELPHI analysis on the measurement of the $b$-quark mass effects 
based on the Durham jet clustering algorithm the impact of the
fragmentation process on the observable $R_3^{b{\ell}}$ was
studied~\cite{Abreu:1997ey} and quantified by adding in quadrature two
different source of errors. The first uncertainty, $\sigma_{tun}$,
was obtained by varying the most relevant parameters of the string
fragmentation model incorporated in JETSET~\cite{jetset} within an 
interval of $\pm 2 \sigma$ from its central value as tuned by
DELPHI~\cite{delphi_tuning} and explained in reference~\cite{Abreu:1997ey}.
The second uncertainty, $\sigma_{mod}$, was the result of analyzing the 
dependence on the fragmentation model itself by comparing the
HERWIG~\cite{herwig} model with JETSET~\cite{jetset}. The difference on
the fragmentation correction factors obtained for each model were
considered as a source of systematic errors. This in fact was the largest
contribution to the total error of the measurement. The
final correction adopted was the average of those two models
and the $fragmentation$ $model$ uncertainty ($\sigma_{mod}$) was
taken to be half of their difference. The total error due to the lack
of knowledge on the hadronization process was expressed as:
\beq
\sigma_{had}(y_c) = \sqrt{\sigma_{tun}^2(y_c)+\sigma_{mod}^2(y_c)} 
\eeq
which at $y_c=0.02$ in the Durham scheme was $\sigma_{had}(y_c)=0.007$ 
\cite{Abreu:1997ey} and its dependence as a function of $y_c$ is shown in 
Fig. \ref{fig:had}. The decision of the measurable $y_c$ interval 
region, $y_c>0.015$ was also connected to the fragmentation correction
which was required to be relatively flat and the four-jet contribution
small ($\leq 2$\%). For comparison purposes, an equivalent analysis has
been performed using the new Cambridge jet reconstruction algorithm and
the results obtained are also presented in Fig. \ref{fig:had}. A larger
flat $y_c$-region is observed in the case of Cambridge with respect to
Durham which can be extended up to $y_c=0.004$ with the four jet
contribution still being small, $\leq$8\%. The total absolute error is
higher for Cambridge than for Durham but the relative sensitivity
to the mass correction is higher for Cambridge at $y_c$ values
around 0.005 than for Durham at $y_c$ values around 0.02. In
Fig.~\ref{fig:had} the difference between the theoretical prediction
of $R_3^{b\ell}$ at LO in terms of the pole mass, $M_b=4.6$ GeV with
respect to that obtained using the running mass $m_b(m_Z)=2.8$ GeV is
also shown. A higher sensitivity to this difference is again found for
the new Cambridge jet algorithm in the valid, flat, $y_c$-region
($y_c>0.004$) thus enabling to test more significantly which of the two 
predictions agrees better with data.

\section{Conclusions} 

We have calculated the next-to-leading order QCD corrections to the
heavy quark three-jet production cross section in $e^+ e^-$-annihilation
as well as the leading order four-jet production cross section using the new
Cambridge jet clustering algorithm. The hadronization corrections
were also estimated.
Comparing with previous studies, this algorithm allows to extend the
analysis into a region of smaller values of the jet resolution parameter,
down to $y_c \approx 0.004$, where the sensitivity to the 
heavy quark mass effects increase.

In particular we have studied in detail the double ratio
$R^{b\ell}_3$ and the differential double ratio $D^{b\ell}_2$.
We have compared the NLO results expressed in
terms of the perturbative pole mass and in terms of the running mass of the
heavy quark at the $m_Z$ scale. We found that the NLO corrections in the case
when the running mass was used are remarkably small.
This is especially true for $R^{b\ell}_3$, where
tree-level expressions in terms of $m_b(m_Z)$ give a very good approximation
to the complete NLO result, which, when expressed in terms of $m_b(m_Z)$,
is almost independent of the renormalization scale. 
In contrast, the calculations done in terms of the pole quark mass have
sizable NLO corrections with a strong renormalization scale dependence.
The hadronization corrections also favor the use
of the Cambridge algorithm with respect to Durham by
keeping a relatively stable uncertainty for small $y_c$.

Summarizing, the results of this paper indicate that a new 
determination of the $b$-quark with the Cambridge jet algorithm will 
improve our present understanding on quark mass effects at LEP/SLC energies.

\acknowledgments

This work has been supported by CICYT under the Grant AEN-96-1718, 
by DGESIC under the Grant PB97-1261 and by the Generalitat Valenciana
under the Grant GV98-01-80.
The work of G.R. has also been supported by INFN and 
E.U. QCDNET contract FMRX-CT98-0194.
The work of M.B. was partly supported by the grant GA\v{C}R-2020506.

\providecommand{\href}[2]{#2}\begingroup\raggedright
\endgroup

\vfill\eject
\mafigura{9 cm}{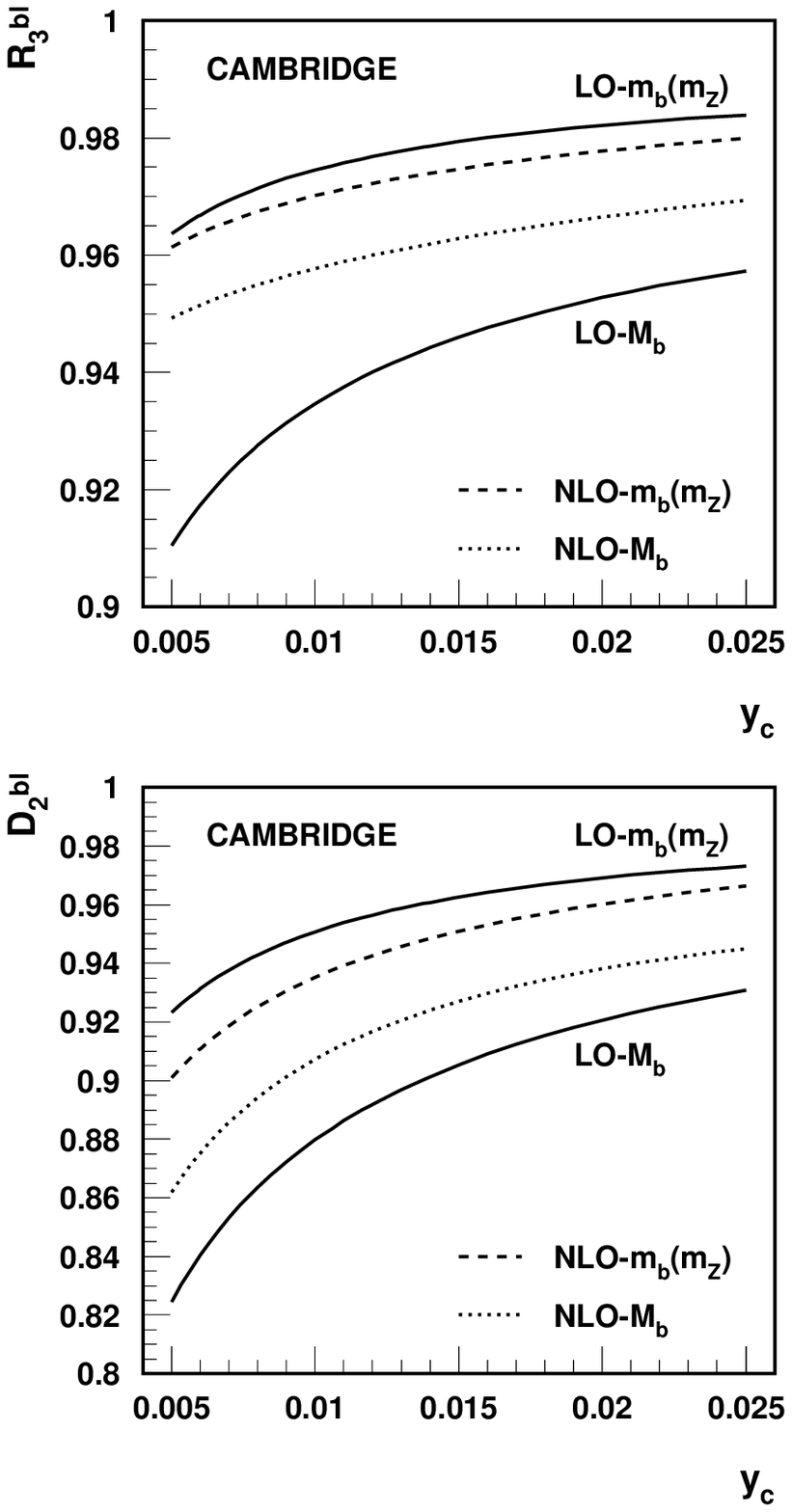}
{The observables $R_3^{b\ell}$ and $D_2^{b\ell}$
as a function of $y_c$ in the Cambridge algorithm at the NLO. 
The dotted lines give the NLO corrected values using~\eq{eq:pole}
for a pole mass of $M_b = 4.6$~GeV.
The dashed lines give the observables at the NLO using~\eq{eq:MS}
for a running mass of $m_b(m_Z) = 2.8$~GeV.
The renormalization scale is fixed to $\mu=m_Z$ and $\alpha_s(m_Z)=0.118$.
For comparison we also plot the LO results for $M_b=4.6$~GeV 
(lower solid lines) and $m_b(m_Z)=2.8$~GeV (upper solid lines).}
{fig:observables}

\mafigura{9 cm}{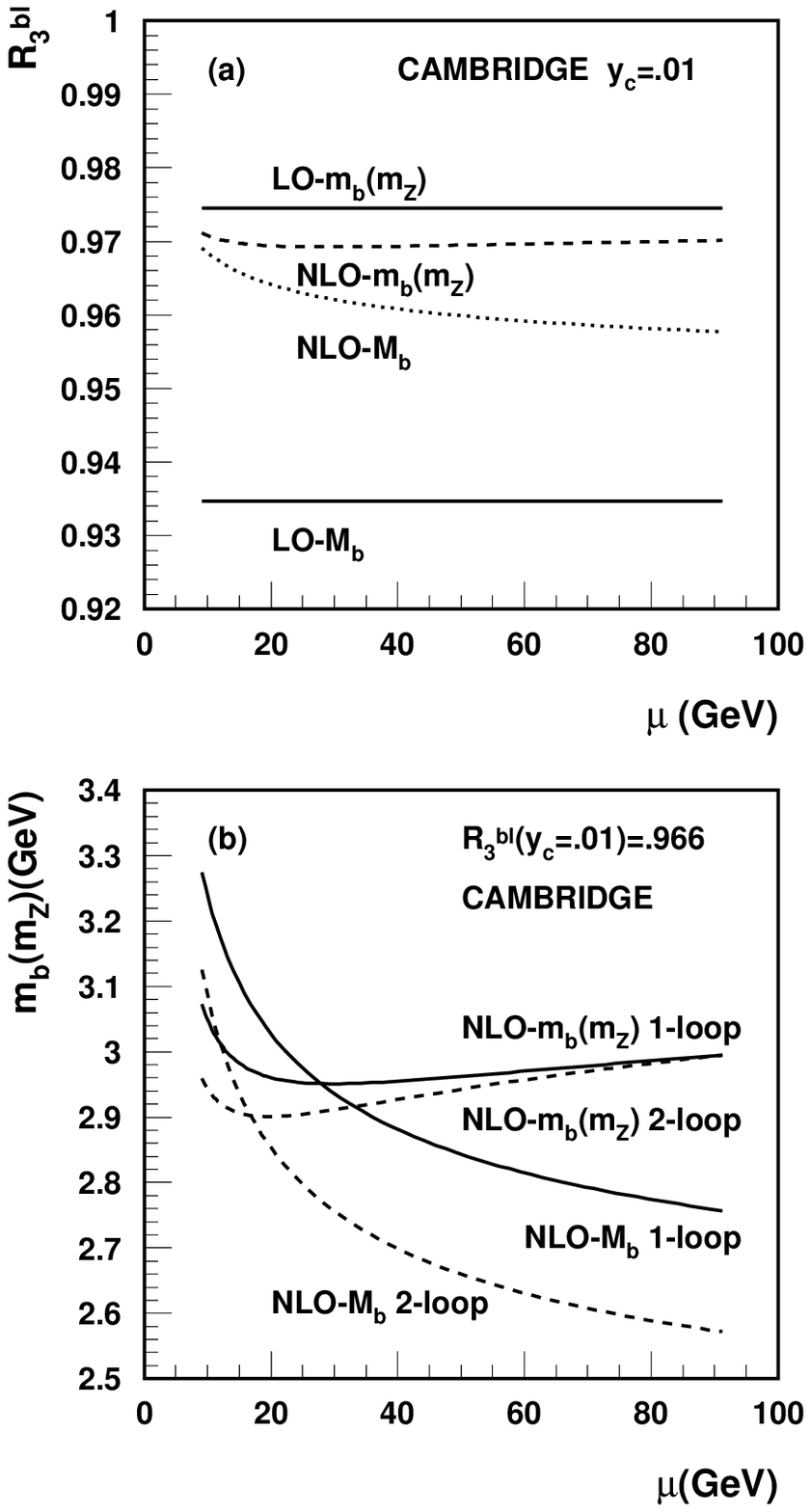}
{a) Renormalization scale dependence for a fixed value of $y_c$. 
Same labels as in Fig.~\ref{fig:observables}.
b) Extracted value of $m_b(m_Z)$ from a fixed value of 
$R_3^{b\ell}$ using either the pole mass expression (NLO-$M_b$)
in~\eq{eq:pole} or the running mass expression (NLO-$m_b(m_Z)$)
in~\eq{eq:MS} as explained in the text.
Solid lines obtained by using one-loop running evolution equations
to connect the results at different scales and dashed lines obtained
by using two-loop expressions.}
{fig:muscales}

\mafigura{10 cm}{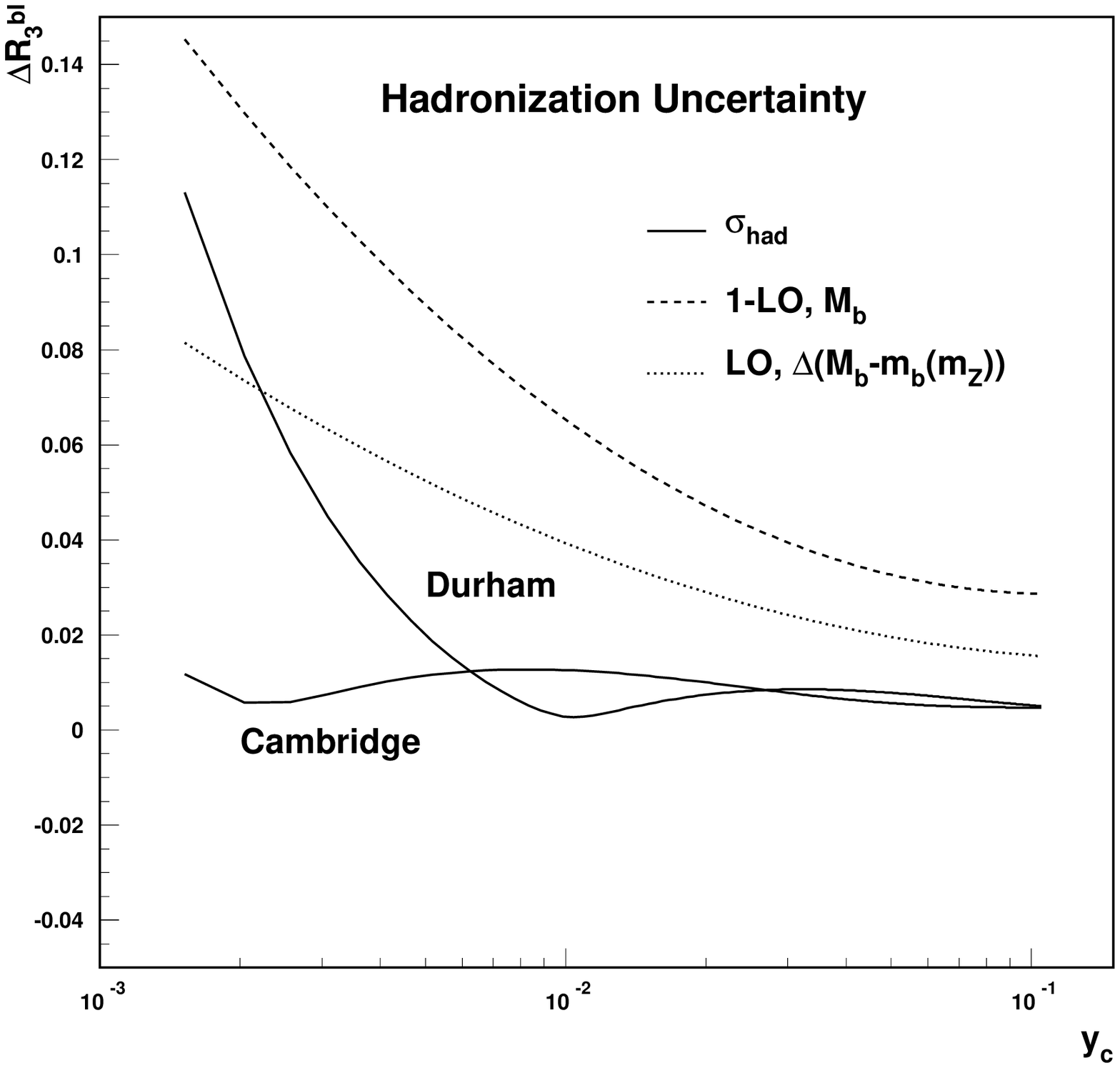}
{Comparison of the hadronization uncertainty ($\sigma_{had}$) obtained
when using either the Cambridge or Durham algorithm. The dashed curve
shows the mass correction at LO for the pole mass, $M_b=4.6$ GeV.
The dotted curve indicates the value of the difference between the LO
predictions for the two mass values, $M_b=4.6$ GeV and $m_b(m_Z)=2.8$ GeV.
The Cambridge algorithm is observed to have a larger stable region on
$y_c$ than Durham reaching at the same time a higher sensitivity to both,
the mass correction and the difference between the two LO predictions.}
{fig:had}


\begin{thebibliography}{10}

\bibitem{Burrows:1998ab}
SLD Coll., P.N. Burrows {\em et.~al.}, {\tt hep-exp/9808017}.

\bibitem{MartiiGracia:1997ak}
S.~Mart\'{\i}~i~Garcia, J.~Fuster, and S.~Cabrera, 
{\em Nucl. Phys. B (Proc. Suppl.)} {\bf 64}, 376 (1998).

\bibitem{Abreu:1997ey}
DELPHI Coll., P.~Abreu {\em et.~al.},
{\em Phys. Lett.} {\bf B418}, 430 (1998).

\bibitem{Rodrigo:1997gy}
G.~Rodrigo, A.~Santamaria, and M.~Bilenky,
{\em Phys. Rev. Lett.} {\bf 79}, 193 (1997). 

\bibitem{Rodrigo:1997gv}
G.~Rodrigo, A.~Santamaria, and M.~Bilenky,
{\tt hep-ph/9905276} to appear in {\em Nucl. Phys.} {\bf B};
{\em J. Phys.} {\bf G25}, 1593 (1999).

\bibitem{Rodrigo:1996gw}
G.~Rodrigo,
{\tt hep-ph/9703359};
{\em Nucl. Phys. B (Proc. Suppl.)} {\bf 54A} 60, (1997).

\bibitem{Brandenburg:1997pu}
A.~Brandenburg and P.~Uwer, {\em Nucl. Phys.} {\bf B515}, 279 (1998).

\bibitem{Bernreuther:1997jn}
W.~Bernreuther, A.~Brandenburg, and P.~Uwer, 
{\em Phys. Rev. Lett.} {\bf 79}, 189 (1997).

\bibitem{Oleari:1997az}
C.~Oleari, ``Next-to-leading order corrections to the production of
heavy flavor jets in $e^+ e^-$ collisions'', {\tt hep-ph/9802431}.

\bibitem{Nason:1997nw}
P.~Nason and C.~Oleari, {\em Nucl. Phys.} {\bf B521}, 237 (1998);
{\em Phys. Lett.} {\bf B407}, 57 (1997).

\bibitem{Catani:1991hj}
S.~Catani, Y.~L. Dokshitser, M.~Olsson, G.~Turnock, and B.~R. Webber, 
  {\em Phys. Lett.} {\bf B269}, 432 (1991).

\bibitem{sld}
SLD Coll., K. Abe {\em et.~al.}, 
{\em Phys. Rev.} {\bf D59}, 012002 (1999). 

\bibitem{Brandenburg:1999nb}
A.~Brandenburg {\it et al.},
``Measurement of the running b-quark mass using $e^+ e^- \to b \bar{b} g$
events'', {\tt hep-ph/9905495}.

\bibitem{opal} 
OPAL Coll., G. Abbiendi {\em et.~al.}, {\tt hep-ex/9904013}.

\bibitem{mbmz}
M.~Jamin and A.~Pich, {\em Nucl. Phys.} {\bf B507}, 334 (1997).
V.~Gim{\'e}nez, G.~Martinelli, and C.~T. Sachrajda, 
{\em Nucl. Phys. B (Proc. Suppl.)} {\bf 53}, 365 (1997);
{\em Phys. Lett.} {\bf B393}, 124 (1997).

\bibitem{Dokshitser:1997in}
Y.~L. Dokshitser, G.~D. Leder, S.~Moretti, and B.~R. Webber, 
{\em JHEP} {\bf 9708}, 001 (1997).

\bibitem{Moretti:1998qx}
S.~Moretti, L.~Lonnblad, and T.~Sj{\"o}strand, 
{\em JHEP} {\bf 9808}, 001 (1998).

\bibitem{Rodrigo:1998vq}
G.~Rodrigo, A.~Santamaria and M.~Bilenky,
{\tt hep-ph/9812433}, {\tt hep-ph/9811465}, {\tt hep-ph/9807489}.

\bibitem{juan98}
DELPHI Coll., J.~Fuster, S.~Cabrera and S.~Mart\'{\i},
contributing paper $\#152$ to ICHEP98 Conf., Vancouver, 1998.

\bibitem{Bentvelsen:1998ug}
S.~Bentvelsen and I.~Meyer, 
{\em Eur. Phys. J.} {\bf C4}, 623 (1998).

\bibitem{Keller:1999tf}
S.~Keller and E.~Laenen,
{\em Phys. Rev.} {\bf D59}, 114004 (1999). 

\bibitem{Bilenkii:1995ad}
M.~Bilenky, G.~Rodrigo, and A.~Santamaria, 
{\em Nucl. Phys.} {\bf B439}, 505 (1995).

\bibitem{Hagiwara:1991dx}
K.~Hagiwara, T.~Kuruma, and Y.~Yamada,
{\em Nucl. Phys.} {\bf B358}, 80 (1991).

\bibitem{Tarrach:1981up}
R.~Tarrach,
{\em Nucl. Phys.} {\bf B183}, 384 (1981).

\bibitem{Rodrigo:1998zd}
G.~Rodrigo, A.~Pich, and A.~Santamaria, 
{\em Phys. Lett.} {\bf B424}, 367 (1998).

%
%
\bibitem{jetset}
T. Sj\"ostrand,
{\em Comp. Phys. Comm.} {\bf 39} 346, (1986).
%
\bibitem{delphi_tuning}
DELPHI Coll., P.~Abreu {\em et.~al.},
{\em Z. Phys.} {\bf C73}, 11 (1996).
%
%
\bibitem{herwig}
G. Marchesini {\em et.~al.},
{\em Comp. Phys. Comm.} {\bf 67} 465, (1992).
%

\end{thebibliography}
\end{document}